\documentclass[fleqn,usenatbib]{mnras}

\usepackage{newtxtext,newtxmath}

\usepackage[T1]{fontenc}
\usepackage{ae,aecompl}

\usepackage{pdfpages}
\usepackage{graphicx}	
\usepackage{amsmath}	
\usepackage{amssymb}	

\title[Correlations Between Diffuse X-rays and the Thermal Sunyaev-Zel'dovich Effect]{Forecasting Angular Cross Correlations Between Diffuse X-ray Emission and the Thermal Sunyaev-Zel'dovich Effect}

\author[Lakey \& Huffenberger]{
Vincent J. Lakey,$^{1}$\thanks{E-mail: vjl12@my.fsu.edu}
Kevin M. Huffenberger,$^{1}$
\\
$^{1}$Department of Physics, Florida State University, 
Tallahassee, FL
32306,
USA
}

\date{\today}

\pubyear{2018}

\begin{document}
\label{firstpage}
\pagerange{\pageref{firstpage}--\pageref{lastpage}}
\maketitle

\begin{abstract}
X-ray emission and the thermal Sunyaev-Zel'dovich distortion to the Cosmic Microwave Background are two important handles on the gas content of the Universe.  The cross-correlation between these effects eliminates noise bias and reduces observational systematic effects.   Using analytic models for the cluster profile, we develop a halo model formalism to study this cross-correlation and apply it to forecast the signal-to-noise of upcoming measurements from eROSITA and the Simons Observatory.  In the soft X-ray band (0.5--2 keV), we forecast a signal-to-noise of 174 for the cross-power spectrum.  Over a wide range of the scales, the X-rays will be signal-dominated, and so sample variance is important.  In particular, non-Gaussian (4-point) contributions to the errors highlight the utility of masking massive clusters.  Masking clusters down to $10^{14}\ M_\odot$ increases the signal-to-noise of the cross-spectrum to 201.   We perform a Fisher Analysis on the fitting coefficients of the Battaglia et al.\ gas profiles and on cosmological parameters.  We find that the cross spectrum is most sensitive to the overall scale of the profiles of pressure and electron density, as well as cosmological parameters $\sigma_8$ and $H_0$, but that the large number of parameters form a degenerate set, which makes extracting the information more challenging.  Our modeling framework is flexible, and in the future, we can easily extend it to forecast the spatial cross-correlations of surveys of X-ray lines available to high-energy-resolution microcalorimetry, to studies of the Warm-Hot Intergalactic Medium, and other effects.
\end{abstract}




\section{Introduction}
\label{sec:intro}
The gas within halos contains untapped information about the baryon content of the Universe. Precise modeling is needed to fully understand how these structures function and evolve, and  to get an accurate census of the baryons in the gas from observations. With observational data, we can compare various modeling schemes---feedback mechanisms, couplings between the gas and star formation, and assembly histories of the gas---to learn about the structure and substructure of the gas. To proceed, we need to model two pieces of information about halos: how the gas is distributed within the individual halos and how the halos are distributed in space. From these, we must compute the quantities that we can observe from the gas.


In this work we examine the cross-correlation of two powerful probes of the gas physics: soft (0.5 - 2.0 keV) X-ray emission and the thermal Sunyaev-Zel'dovich (tSZ) effect.  Using them in cross-correlation can help to avoid biases from systematics in a single observable.




In X-rays,  the gas emits through bremsstrahlung \citep{Bohringer2000,Ebeling2000,Ebeling2001} and line transitions \citep{Sarazin1977}. The observed X-ray emissivity of the gas scales like the square of the electron density of the gas, so it is especially sensitive to the gas's core structure within a virial radius.  This is true for both the continuum and collisionally-ionized line emission in the gas.
Soft X-rays are interesting because they may harbor signals from the missing baryons in the Warm-Hot Intergalactic Medium (WHIM).  Multiple additional components pollute the diffuse X-ray background, including signals from AGN, the Milky Way, the Local Hot Bubble, and charge exchange in the Solar wind. (For an overview, see \mbox{\citet{Galeazzi2009}}.) This contamination overlaps the continuum emission as well as key lines (specifically Oxygen).

The same gas also distorts the spectrum of Cosmic Microwave Background (CMB) photons passing through it.  The tSZ effect \citep{Sunyaev1969,Sunyaev1972} inverse-Compton scatters photons to higher energy, leading to a photon deficit at frequencies lower than 217 GHz when compared to lines-of-sight with no interaction, and an excess above 217 GHz.  We observed the effect as an effective temperature decrement or increment in the CMB at those frequency ranges.  The temperature signal is proportional to the line-of-sight integral of the gas pressure. In comparison to X-ray emission, the SZ probes more of the lower density gas, and the signal is less concentrated toward the center of the halo.  Components that contaminate the X-rays, like AGN and the Milky Way, also produce emission in the microwaves, but the tSZ effect has a unique frequency dependence that changes sign across microwave frequencies, unlike these contaminants.  The tSZ effect is well observed, and is a main avenue to find galaxy clusters, particularly at high redshifts (e.g.  \cite{Carlstrom2002, Hand2011, Sehgal2013, Hasselfield2013, Lagana2013, Bleem2014, Liu2015, Hilton2018}).




In order to study the cross correlation, we seek instruments that measure the effects over a large, overlapping sky area. We examine eROSITA \citep{Merloni2012} for X-ray emission and the Simons Observatory \citep{SO2018} for tSZ.  eROSITA is the primary instrument aboard the Russian \textit{Spectrum-Roentgen-Gamma} satellite set to launch in 2019. This mission is a successor to ROSAT \citep{Voges1999} and intends to survey the entire X-ray sky in the soft (0.5 - 2.0 keV) and hard (2.0 - 10.0 keV) X-ray bands. For our purposes, we focus on the soft X-ray band, where eROSITA is 20 times more sensitive than ROSAT. For tSZ, we look to Simons Observatory, a next-generation collection of CMB telescopes based in Chile, which will survey a large area of the Southern sky.  The relevant data product from SO is a sky-component-separated Compton-$y$ map, similar to those used in \cite{Planck2013, PlanckXXII2015, Hill2014, Khatri2016}. Combined, these two data sets will provide the best combination of sensitivity and large sky coverage available by the mid 2020s.  

Here we compute angular power spectra using a halo model approach.
The halo model is a standard technique for computing clustering statistics of signals from halos \citep[for a review, see][]{CooraySheth}. It treats halos as idealized spheres, with matter and gas radial profiles that vary as as a function of redshift and halo mass.  The halo mass function provides the number counts of such halos. Integrating over the halo population, the total angular power spectrum gives us a measurement of the net correlations between the observables  in harmonic space.


On a halo by halo basis, we model both X-ray and tSZ effects in real space, then Fourier transform them to compute the angular power spectrum. 
This same cross-correlation has been studied  by similar modeling \citep{Hurier2014, Singh2015} and detected to $28\sigma$ significance with current ROSAT and Planck data \citep{Hurier2015}.  The goal of our work is to update the modeling, provide an independent examination of these effects, and forecast the constraints that future experiment will allow us to set on cosmological and gas parameters.


This paper is organized as follows: in Section \ref{sec:methods} we describe the gas modeling, X-ray emission, the tSZ effect, and the halo model approach to compute angular power spectra. In Section \ref{sec:results} we discuss the results of angular power spectra and Fisher analysis. In Section \ref{sec:discussion} we discuss implications of our analysis as well as future investigations.


\section{Methods}
\label{sec:methods}
For our purpose, the gas in halos have two fundamental quantities that we need to model: pressure and electron density. From these two quantities, we can calculate the gas emission in X-rays and the tSZ effect's change to the CMB's temperature.  For both effects, we can integrate over the distribution of halos to predict the angular power spectrum in a halo model approach. We can make predictions of the contribution to the mean X-ray background and the tSZ auto-power spectrum to check the validity of our procedure against previous works and calculate the  angular cross-power spectra for specific halo mass ranges.  With noise estimates of upcoming experiments, we can forecast constraints on cosmological and gas fitting coefficients.

\subsection{Halo Gas Model}
\label{sec:gasmodel}
Our analysis depends crucially on the pressure and distribution of electrons in the gas. The gas is non-uniform and interacts with by the environment in a myriad of ways: feedback from Active Galactic Nuclei (AGN) and supernovae, radiative cooling, star formation, galactic winds, and cosmic rays \citep{Battaglia2011}. All of these effects must be considered in order to get a full picture of the gas. We model the gas with fitting functions to hydrodynamical simulations that include these effects. \citet{Battaglia2011} uses empirical fits of hydrodynamical simulations to obtain a robust electron pressure profile for the gas as a function of distance from the halo center: 
\begin{equation}
P(r) = P_0 (x/x_c)^\gamma [1+(x/x_c)^\alpha]^{-\beta},
\end{equation}
where $x=r/R_{200}$ and $R_{200}$ is the radius at which the average halo density is 200 times the critical density.  Due to a parameter degeneracy described in \cite{Battaglia2011}, we fix $\alpha = 1.0$ and $\gamma = -0.3$ for our analysis (as they did).  Each of these model parameters have an overall scale, mass dependence, and redshift dependence.  For a generic parameter $A \in \{P_0, x_c, \beta\}$ these are given by:
\begin{equation}
A = A_0 \left( \frac{M_{200}}{10^{14}M_\odot}\right)^{\alpha_m} (1+z)^{\alpha_z}.
\label{equation:fit}
\end{equation}
where $A_0$ is the scale coefficient, $\alpha_m$ is the mass dependence coefficient, and $\alpha_z$ is the redshift dependence coefficient.
All of these gas fit coefficients are found in Table \ref{table:gasparams}. From \cite{Battaglia2016a}, we use a similar fitting function for the gas density: 
\begin{equation}
\rho_{\textnormal{gas}}(r) = \rho_0 (x/x_c)^{\gamma} [1 + (x/x_c)^\alpha]^{-({\beta+\gamma})/{\alpha}}.
\end{equation}
Following them we fix $x_c = 0.5$ and $\gamma = -0.2$, again due to parameter degeneracies, and the other parameters have fit coefficients that scale like Equation \ref{equation:fit}. From the gas density, the electron density is given by:
\begin{equation}
n_e(r) = \frac{\rho_{\textnormal{gas}}(r) x_e  X_H  (1-f_\star)f_c}{m_p},
\end{equation}
where $X_H = 0.76$ is the primordial hydrogen mass fraction, $x_e = {(X_H + 1)}/{2X_H}$ is the electron fraction, $f_\star = 0.02$ is the stellar mass fraction of the halo, and $f_c = 1.0$ is the correction for the baryon depletion \citep{Battaglia2011}. We get the electron temperature profile $T_e(r)$ from the pressure and density using the ideal gas law.

Since the gas is inhomogeneously distributed on small scales, we define its clumping factor,
\begin{equation}
 C_{2\rho} = \frac{\langle n_e^2 \rangle}{\langle n_e \rangle^2},
\end{equation} \citep{Battaglia2014} as a correction on the squared density of the gas given by:
\begin{equation}
C_{2\rho}(r) = 1 + (x/x_c)^{\beta} [1 + (x/x_c)]^{\gamma - \beta}.
\end{equation}
For a parameter $q \in \{x_c, \beta, \gamma\}$, the scaling with mass and redshift is given by:
\begin{equation}
q(M_{200}) = q_1 \left( \frac{M_{200}}{10^{14} M_\odot} \right)^{q_2},
\end{equation}
where the mass dependence is explicitly stated but the redshift dependence is implicit inside of the definition of $M_{200}$.  These parameters are likewise summarized in Table~\ref{table:gasparams}.

Previous works \citep{Hurier2014, Hurier2015, Singh2015} used a generalized Navarro-Frenk-White profile to model the gas pressure.  In \citet{Hurier2015}, they used a polytropic equation of state to compute the electron density whereas in \cite{Singh2015} they approximated the temperature of the gas to be uniform at the virial temperature. Their methodology has the benefit of having a small number of free parameters related to gas modeling. For our work, we chose a more complex model to allow for a more complete parameterization of the gas. This comes at the expense of having a much larger set of parameters in our analysis, which makes them more difficult to independently constrain.

\begin{table}
\centering
\begin{tabular}{||c c c c c||} 
 \hline
 Quantity & Param.  & \multicolumn{3}{c|}{Fitting Coefficients}{} \\ [0.5ex] 
 \hline 
     & & Scale & Mass dep. & Redshift dep. \\ 
 \hline \hline
 $P$ & & $A_0$ & $\alpha_m$ & $\alpha_z$ \\
 \hline
 & $P_0$ & $18.1$ & $0.154$ & $-0.758$ \\
 & $x_c$ & $0.497$ & $-0.00865$ & $0.731$ \\
 & $\beta$ & $4.35$ & $0.0393$ & $0.415$ \\
 \hline
 $\rho_{\rm gas}$ & & $A_0$ & $\alpha_m$ & $\alpha_z$ \\
 \hline
 & $\rho_0$  & $4 \times 10^3$ & $0.29$ & $-0.66$ \\
 & $\alpha$ & $0.88$ & $-0.03$ & $0.19$ \\
 & $\beta$ & $3.83$ & $0.04$ & $-0.025$ \\
 \hline
 $C_{2\rho}$ & & $q_1$ & $q_2$ & \\
 \hline
& $x_c$  & $9.91 \times 10^5$ & $-4.87$ & \\ 
& $\beta$ & $0.185$ & $0.547$ & \\
& $\gamma$ & $1.19 \times 10^6$ & $-4.86$ & \\
[1ex] 
 \hline
\end{tabular}
\caption{Fitting coefficients from Battaglia profiles for the pressure, gas density, and gas clumping factor.}
\label{table:gasparams}
\end{table}



\subsection{X-ray Emission}
\label{sec:xray}
Due to its temperature, the gas in $M >10^{13} M_\odot$ halos emits in the X-ray band.
The amount of emission a halo produces is given by its emissivity, $\epsilon(T_e)$, the energy emitted per time per volume per number density squared.  We use the Astrophysical Plasma Emission Code \citep[APEC,][]{Smith2001} to calculate the X-ray emissivity of the gas within our halos.  APEC allows us to calculate the emissivity on a process by process basis and is split into two major components: continuum emission and line emission. Continuum emission includes all the processes that occur through radiative recombination---when electrons collide and recombine with ions---and bremsstrahlung, the braking of electrons around charged particles.  Line emission includes electronic transitions to lower energy states of elements in the hot plasma and are calculated through equilibrium level populations, transition rates, and elemental abundances in the gas. We sum the continuum and line emissivities and integrate them over the emitted frame energy band.
Because of the $K$-correction, this band is redshift dependent and corresponds to the energy that will redshift into the 0.5-2.0 keV band in the observer frame.  This yields the total rest-frame emissivity for the gas in the energy band of interest:
\begin{equation}
\epsilon(T_e) = \int_{(1+z)E_{\rm min}}^{(1+z)E_{\rm max}} \left( \frac{d\epsilon}{dE} \right)_{\textnormal{APEC}} dE,
\end{equation}
where $\left( d\epsilon / dE \right)_{\textnormal{APEC}}$ is the sum of the continuum and line emissivities provided by APEC.  In our application we assume solar metallicities, but here we are not very sensitive to metallicity because the continuum dominates our relatively wide band. From these emissivities, we can calculate
the differential luminosity in the emitter frame:
\begin{equation}
    dL = \epsilon(T_e) C_{2\rho}n_en_H dV
\end{equation}
where $n_H$ is the hydrogen number density in the halo and the proper volume element is $dV = dr_{\rm los} d_A^2 d\Omega$, for  line-of-sight proper distance $r_{\rm{los}}$ and  angular diameter distance $d_A$. 

Thus we obtain the total X-ray surface brightness: 
\begin{equation}
X = \int dr_{\rm{los}} \frac{\epsilon(T_e(r_{\rm los})) C_{2\rho}(r_{\rm{los}})n_e(r_{\rm{los}})n_H(r_{\rm{los}})}{4\pi(1+z)^{4}}.
\end{equation}
The factor $(1+z)^4$ comes from the ratio of $d_A^2/d_L^2$ where $d_L$ is the luminosity distance, and this ratio accounts for the volume element, the distance dependence, and the redshifting of the X-ray emission.
We define the 3-dimensional X-ray emission profile as the integrand of the surface brightness integral, but measured for each halo from the halo center:
\begin{equation}
x_{3D}(r) = \frac{\epsilon(T_e(r)) C_{2\rho}(r) n_e(r)n_H(r)}{4\pi(1+z)^{4}}.
\end{equation}
  The Fourier transform of the halo profile is an important ingredient in the computation of the power spectrum.  For the X-ray emission, the Fourier Transform of the profile, $x_l$, is:
\begin{equation}
x_l = \frac{4\pi}{{d_A}^2} \int dr r^2 \frac{\sin(lr/d_A)}{lr/d_A} x_{3D}(r),
\end{equation}
where $d_A$ is the angular diameter distance and multipole $l$ is the angular wave number.  For this work, we take $l=40$ to $l=7930$ with $\Delta l = 50$ and evaluate the integral from $r_{\textnormal{min}} = 10^{-6}$ Mpc to $r_{\textnormal{max}} = 3R_{200}$ in $10^{4}$ logarithmic-spaced bins.

The approaches in previous works were similar, but differ in the details.  In \cite{Singh2015}, they used a cooling function approach to model the emissivity of X-rays. This approach makes sense due to their choice of a constant temperature for the halo and allowed their cooling function to be only a function of the gas metallicity. In \cite{Hurier2014}, they used a MEKAL \citep{MEWE1985} model for the X-ray emission of the gas and allowed the emission to scale with gas metallicity. This allowed them to have a radius, mass, metallicity, and redshift dependent function for the X-ray emission of the gas. In our work, we chose APEC for its versatility, allowing us to examine distinct parts of the emission.  This lets us calculate many different emission scenarios for this and future works.

\subsection{Mean X-ray Background}
\label{sec:xrbkg}
The mean X-ray background is the summed effect of all X-rays across the sky with unknown origin. Our model makes a concrete prediction for the contribution to the X-ray background tied to unresolved halos. This does not account for AGN emission or Galactic sources. Using our halo model formalism, the mean X-ray background from halos is given by:
\begin{equation}
\langle x \rangle = \frac{1}{4\pi}\int_{0}^{z_{\rm max}} dz \frac{dV}{dz} \int_{M_{\rm min}}^{M_{\rm max}} dM \frac{dn}{dM} {x_{l=0}}(M,z),
 \end{equation}
where $x_{l=0}$ is the Fourier transform of $x_{3D}$ taken only for the $l=0$ monopole.
For the halo abundance as a function of mass and redshift we use the halo mass function of \cite{Tinker2008},  a fit to $N$-body simulations.  It provides $dn/dM$, which is the differential comoving number density of halos of a given mass.  We compute the mass function from the Python module \textit{hmf} \citep{Murray2013} and convert the halo mass to $M_{200c}$ to use with the Battaglia et al.\ fitting functions.   The differential comoving volume per steradian  per redshift interval, $dV/dz$, is cosmology dependent, and we compute it with the Python code \textit{astropy} \citep{Astropy1, Astropy2}.


\subsection{Thermal Sunyaev-Zel'dovich Effect}
\label{sec:tsz}
The tSZ effect causes a change in inferred temperature of the CMB photons, $\Delta T_{\textnormal{tSZ}}$:
\begin{equation}
\frac{\Delta T_{\textnormal{tSZ}}}{T_{\textnormal{CMB}}} = g(\nu) y, 
\end{equation}
where $g(\nu) = x \coth (x/2) - 4$ is the frequency dependence of this effect with $x = {h\nu}/{k_B T_{CMB}}$ and $y$ is the Compton $y$ parameter. The Compton $y$ parameter is a line of sight integral of electron pressure of the gas that causes the scattering:
\begin{equation}
y = \int dr_{\rm{los}} \frac{\sigma_T}{m_ec^2} P(r_{\rm{los}}),
\end{equation}
where $\sigma_T$ is the Thompson cross section of the scattering. From the Compton $y$ parameter, we extract the 3D tSZ profile of the halo: 
\begin{equation}
y_{3D}(r) = \frac{\sigma_T}{m_ec^2} P(r).
\end{equation}
We then take the Fourier transform of the profile:
\begin{equation}
y_l = \frac{4\pi}{{d_A}^2} \int dr r^2 \frac{\sin(lr/d_A)}{lr/d_A} y_{3D}(r).
\end{equation}


\subsection{Angular Power Spectrum}
\label{sec:cl}

We can compute the angular power spectrum with the halo model formalism \citep[e.g.][]{Cooray2001}.  First we compute the one-halo term, due to correlations of effects between two points within a single halo:
\begin{equation}
C_{l}^{ss',1h} = \int_{0}^{z_{\rm max}} dz \frac{dV}{dz} \int_{M_{\rm min}}^{M_{\rm max}} dM \frac{dn}{dM} [s_{l}(M,z) \times s'_{l}(M,z)],
\end{equation}
where  $s$ and $s'$ are signals that are given by either $x_l$ or $g(\nu)y_l$. 

In this work we use $z_{\textnormal{max}} = 5$, $M_{\textnormal{min}} = 10^{11} M_{\odot}$, $M_{\textnormal{max}}$ varies between $10^{13}-10^{15} M_{\odot}$. Lowering $M_{\textnormal{min}}$ to $M = 10^{10} M_\odot$ only provides a sub-percent increase to the power spectra for the highest maximum mass and only a three percent correction to the lowest. For halos of $M > 10^{15} M_\odot$, the halo population is so small that there is no significant contribution for our work. The redshifts binning scheme uses 50 logarithmic bins between $z=10^{-3}$  and $z=1$ (to resolve the effect of the steeply changing comoving volume and X-ray flux) and 50 linear bins between redshifts $z=1$ and $z=5$ (to capture the SZ effect for distant halos). The masses are binned logarithmically in 200 bins.  For concreteness we consider the tSZ signal at 150 GHz, where it is negative, and show the absolute value of the correlation.

Correlations between effects in two separate halos are given by the two-halo term:
\begin{equation}
C_l^{ss',2h} = \int_{0}^{z_{\rm max}} dz \frac{dV}{dz} P_{m}(k = {l}/{r}, z) [W_l^s(z) \times W_l^{s'}(z)],
\end{equation}
where $P_m$ is the matter power spectrum and $W_l(z)$ is the window function given by:
\begin{equation}
W_l^s(z) = \int_{M_{\rm min}}^{M_{\rm max}} 
dM \frac{dn}{dM}(M,Z) b(M,z) s_l(M,z),
\end{equation}
where $b(M,z)$ is the halo bias.  We use \textit{hmf}'s implementation of both the matter power spectrum and the halo bias. The total angular power spectrum is given by the summation of the one- and two-halo correlation terms:
\begin{equation}
C^{ss'}_l = C^{ss',1h}_l + C^{ss', 2h}_l.
\end{equation}

\subsection{Forecasting Future Experiments}
\label{sec:errors}
With our model in hand, we must now investigate the detectability of these signals. The eROSITA mission is an upcoming X-ray satellite-based all-sky surveyor that will detect soft emission with a much higher sensitivity than its predecessors. The noise power spectrum for eROSITA soft X-ray emission is given by the eROSITA Science Book \citep{Merloni2012}:
\begin{equation}
N_l^{xx} = N_{\textnormal{bkg}} \exp \left( \frac{l(l+1)\theta^2_{\textnormal{FWHM}}}{8 \ln 2} \right),
\end{equation}
where $N_{\textnormal{bkg}}$ is given by:
\begin{equation}
N_{\textnormal{bkg}} = \frac{\langle s \rangle E_{\textnormal{photon}}}{t_{\textnormal{exp}}A_{\textnormal{eff}}},
\end{equation}
where $E_{\textnormal{photon}}$ is the average photon energy (taken to be 1 keV), $t_{\textnormal{exp}}$ is the exposure time for the survey, $A_{\textnormal{eff}}$ is the effective area of the instrument, $\theta_{\textnormal{FWHM}}$ is the full width half maximum beam width and $\langle s \rangle$ is the simulated background photon count rate from eROSITA. These experimental values for eROSITA, converted to the units we use here, are shown in Table \ref{table:1}. 

We take the noise power spectrum for the tSZ effect from the forecast in  \cite{SO2018}.\footnote{Simons Observatory noise curves publicly available at:  https://simonsobservatory.org}  That estimate is an internal linear combination of the ``baseline''-model predicted noise power spectra from the Simons Observatory at 27, 39, 93, 145, 225, and 280 GHz and Planck at 30, 44, 70, 100, 143, 217, and 353 GHz. The combination of frequency channels minimizes the variance for each multipole. 

From the halo model formalism, we can also calculate the bin--bin covariance of the cross-spectrum \citep{Knox1995, Cooray2001,Komatsu2002}:
\begin{equation}
\sigma^2_{ll'} = f_{\textnormal{sky}}^{-1} \left[2 (\Delta C_l^{xy})^2 \delta_{ll'} + \frac{T_{ll'}^{xy}}{4\pi} \right],
\label{sigmasquared}
\end{equation}
where $f_{\textnormal{sky}}$ is the fraction of the sky that is overlapping between the two data sets and $\Delta C_l^{xy}$ is given by :
\begin{equation}
(\Delta C_l^{xy})^2 = \frac{1}{(2l+1)\Delta l} [(C_l^{xx} + N_l^{xx})(C_l^{yy} + N_l^{yy}) + (C_l^{xy})^2],
\end{equation}
where $\Delta l$ is the bin size. The trispectrum $T_{ll'}^{xy}$ is defined as: 
\begin{equation}
T_{ll'}^{xy} = g(\nu)^2 \int_{0}^{z_{max}} dz \frac{dV}{dz} \int_{M_{min}}^{M_{max}} dM \frac{dn}{dM} [x_{l}
\times y_{l} \times x_{l'} \times y_{l'}],
\end{equation}
and gives the 4-point contribution to the power spectrum covariance that arises because the signals are non-Gaussian.
In order to estimate the total signal-to-noise from the cross correlation, we estimate the variance of an amplitude, $A$, fit to a known spectral shape:
\begin{equation}
C_l^{xy} = A C_{0,l}^{xy}
\end{equation}
where $A = 1$ is the true value of $A$. We then look at the variance on $A$:
\begin{equation}
\textnormal{var}(A) = \bigg[ \sum_{ll'} \frac{\partial C_l}{\partial A} \sigma_{ll'}^{-2} \frac{\partial C_l}{\partial A} \bigg]^{-1} = \sigma_A^2
\end{equation}
and from that calculate the total signal to noise ratio (SNR) as $\textnormal{SNR} = A/\sigma_A$.

\begin{table}
\centering
\begin{tabular}{||c c||} 
 \hline
 Parameter & Value \\ [0.5ex] 
 \hline\hline
 $f_\textnormal{sky}$ & 0.4 \\
 \hline
\multicolumn{2}{|c|}{eROSITA-like survey} \\
 \hline
 $t_{\textnormal{exp}}$ & 2 ks \\ 
 $A_{\textnormal{eff}}$ & 1500 $\textnormal{cm}^2$ \\
 $\theta_{\textnormal{FWHM}}$ & 20 arcsec \\
 $\langle s \rangle$ & $158.717 \ \textnormal{keV} / \textnormal{cm}^2 / \textnormal{s} / \textnormal{sr}$ \\
 \hline
\end{tabular}
\caption{Sky overlap and survey parameters for an upcoming eROSITA-like data set.}
\label{table:1}
\end{table}

\section{Results}
\label{sec:results}

We check our model by comparing to measurements of the tSZ power spectrum and the mean diffuse X-ray background.  Then we present our prediction for the cross-power spectrum and forecast measurement errors on it.  Finally, we explore the gas and cosmological model parameters to which such a measurement would be sensitive.

\subsection{Robustness checks}
\label{sec:power}
\begin{figure}
\includegraphics[width=\linewidth]{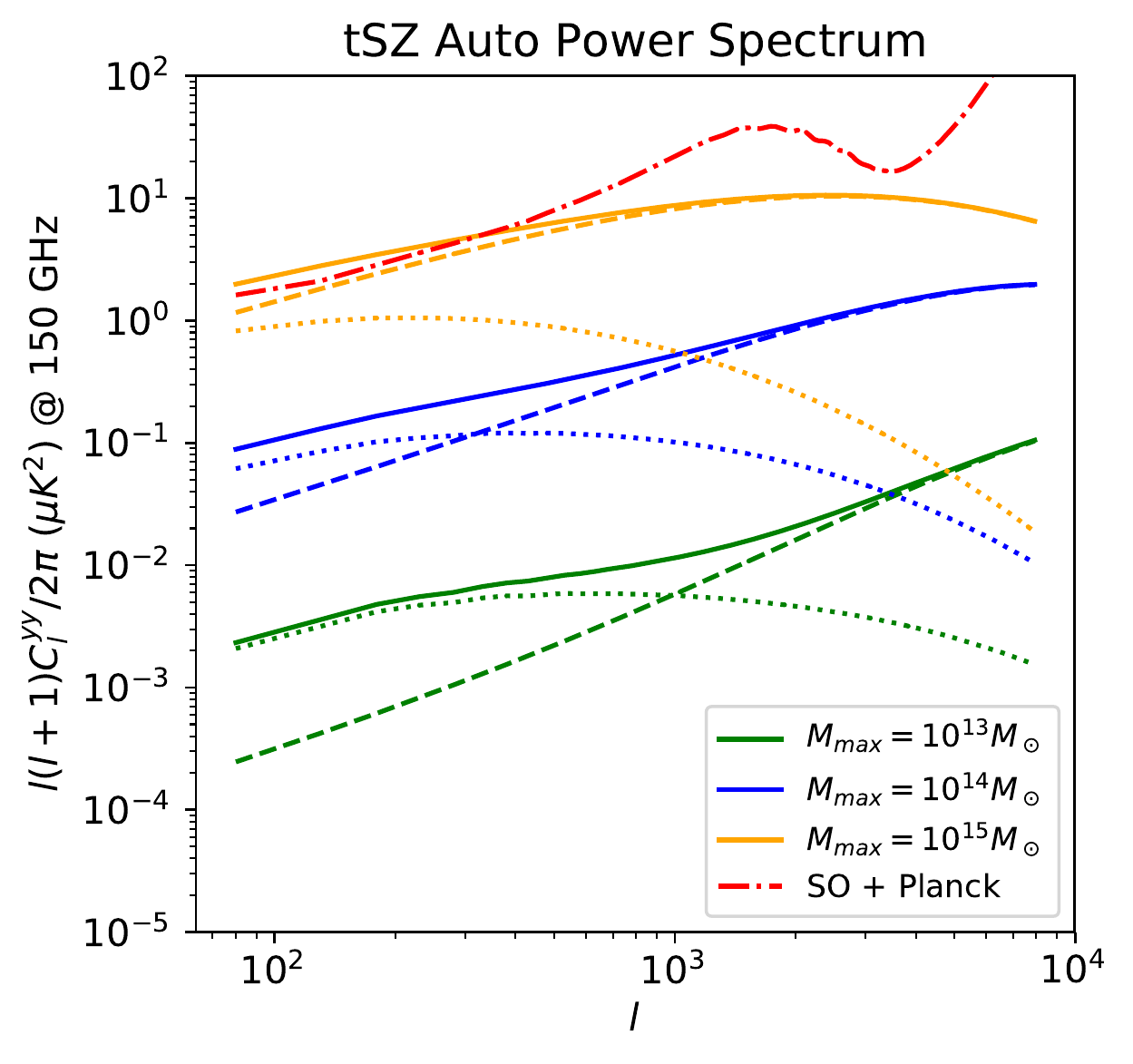}
\includegraphics[width=\linewidth]{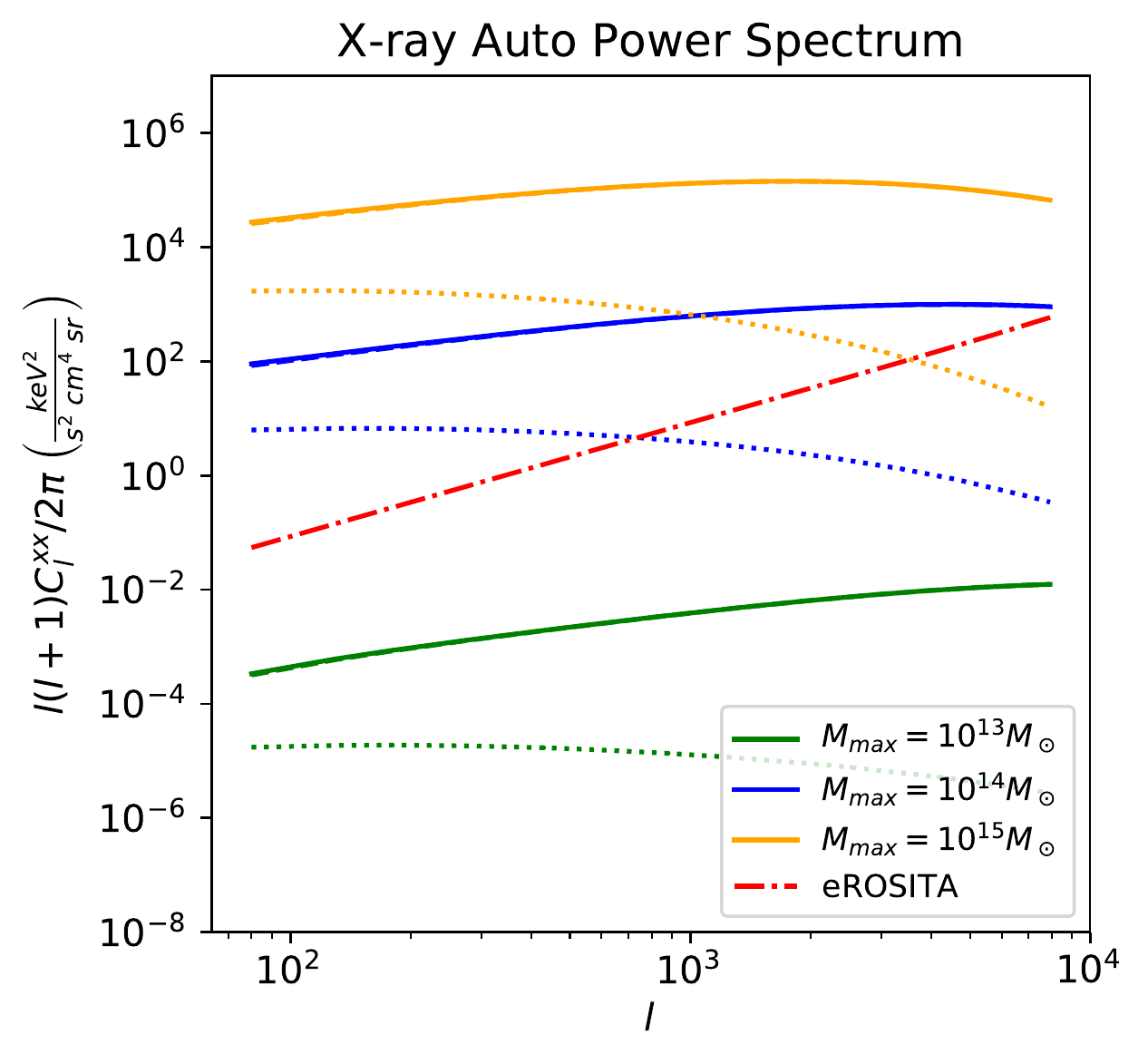}
\caption{Above: tSZ auto-power spectrum for $M < 10^{15} M_\odot$ (orange), $M< 10^{14} M_\odot$ (blue) and $M < 10^{13} M_\odot$ (green) halos with SO-like noise curves (red). Dashed curves are the one-halo terms and dotted curves are the two-halo terms. 
Although the Compton-$y$ noise level is built from multiple channels of the Simons Observatory and Planck, we scale it to an equivalent tSZ power spectrum at 150 GHz.
Below: X-ray auto-power spectrum with eROSITA photon shot noise (red). The one-halo terms dominate, and so the dashed lines are mostly covered by the solid lines.  As mass increases, the correlation shifts left in $l$ due to the size of the halos increasing.}
\label{CxxCyy}
\end{figure}

We show the tSZ and X-ray power spectra in Figure \ref{CxxCyy} for various maximum mass cuts, along with the experimental noise models.  As more massive halos are included, the peak of the power spectrum shifts to lower $l$ due to the average size of objects increasing. Our analysis matches \cite{Battaglia2011} and is in agreement with the \citet{Planck2013} measurement of the Compton $y$ map auto-spectra, within 1$\sigma$ for most multipoles.

\begin{figure}
\includegraphics[width=\linewidth]{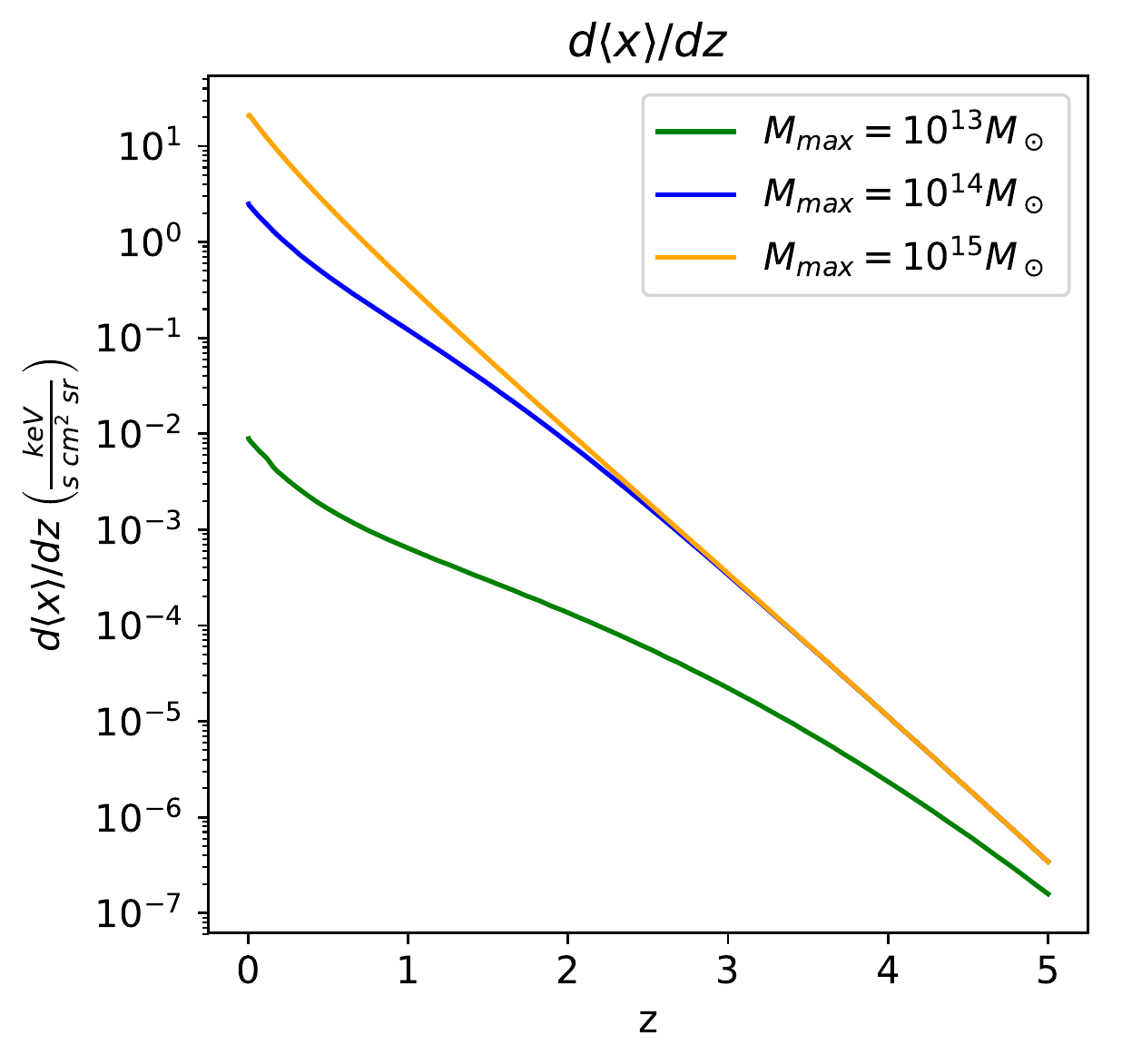}
\includegraphics[width=\linewidth]{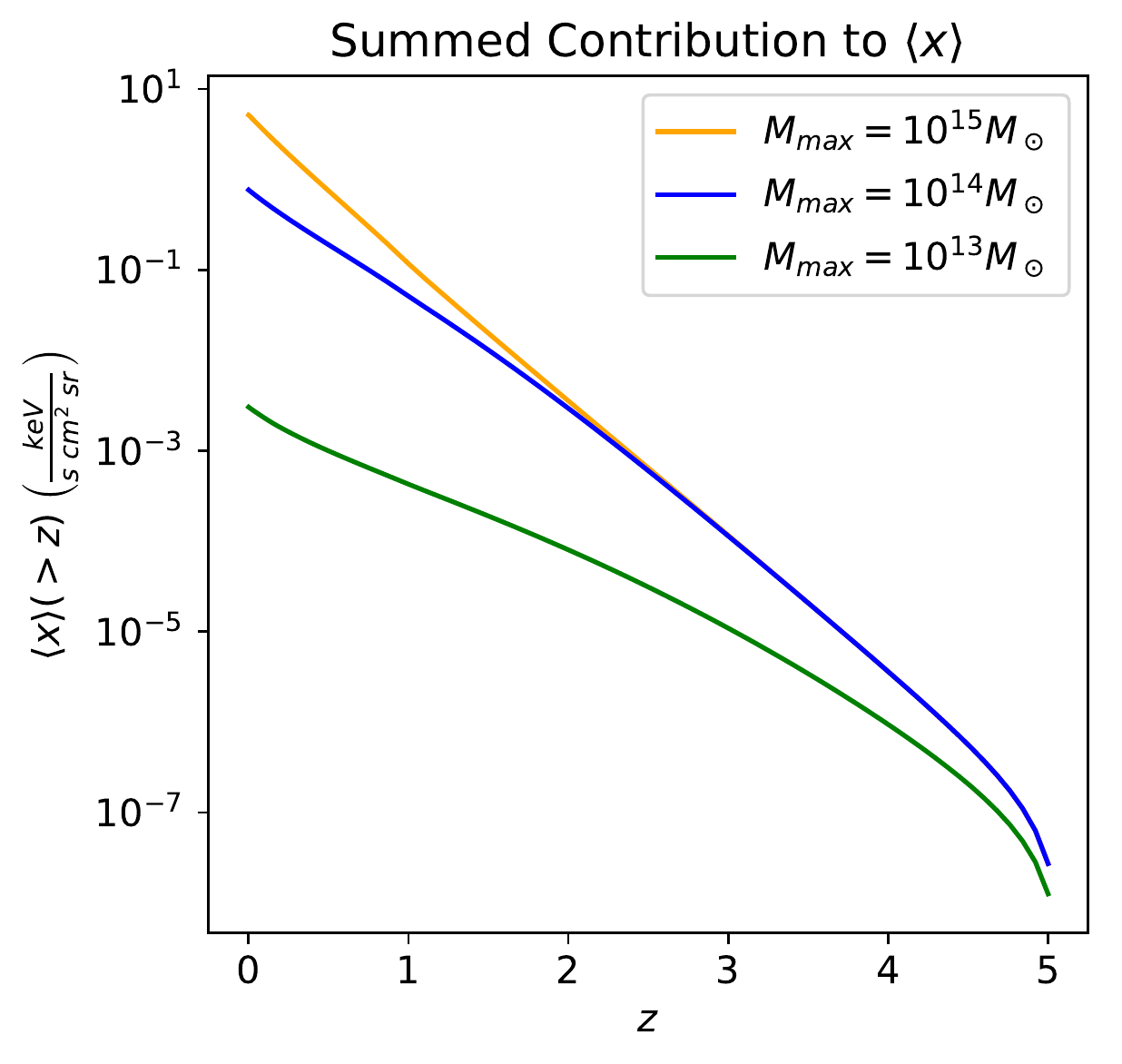}
\caption{Top: Contribution to the mean X-ray background $d\langle x \rangle /dz$ versus redshift $z$ for each mass cut. The top two mass ranges converge around $z=2$ due to the relative number of clusters falling off. All three mass ranges converge at $z>5$. Bottom: Total summed contribution to the mean X-ray background for each mass cut. The total mean X-ray background contribution is at $z=0$.}
\label{MeanS}
\end{figure}

We also examined the mean X-ray background contribution from halos.  Figure \ref{MeanS} shows our results for the mean X-ray background in two ways: the differential X-ray background contribution per redshift interval and the cumulative sum of the background over time. For low redshifts, the signal is dominated heavily by clusters and groups with $M = 10^{14}$--$10^{15} M_\odot$.  The contribution from halos with $M<10^{13} \ M_\odot$ is only $\sim 10^{-3}$ of the  total halo contribution at $z=0$. 
At higher redshift the abundance of massive halos drops, and so the contribution from smaller halos become relatively more important.  We did not tune any parameter of the X-ray model to achieve a particular outcome for the mean X-ray background; we simply grafted the APEC emission model onto the Battaglia et al.\ gas models and integrated over the halo abundance.  


The cumulative redshift integration of X-ray flux yields the total X-ray background that halos (up to $M = 10^{15} M_\odot$) produce in our energy band.
Some 50 percent of the signal arises from $z < 0.14$ and 90 percent is from $z < 0.5$.
At $z = 0$, our summed total contribution is 
$5.1 \ \textnormal{keV} / \textnormal{cm}^2 / \textnormal{s} / \textnormal{sr}$ in the 0.5--2 keV band for $M < 10^{15} M_\odot$ halos. For $M < 10^{14} M_\odot$ halos, this number is  $0.73 \ \textnormal{keV} / \textnormal{cm}^2 / \textnormal{s} / \textnormal{sr}$. Finally, for $M < 10^{13} M_\odot$ halos, we find $3.0 \times 10^{-3} \ \textnormal{keV} / \textnormal{cm}^2 / \textnormal{s} / \textnormal{sr}$. 

A fair comparison of these values to the literature is tricky.  \citet{HM06} fit an extragalactic component to spectral measurements of CHANDRA Deep Field North and Deep Field South observations that had resolved sources masked.  The component was a power law in energy, modified by Galactic absorption.  In the 0.5--2.0 keV band, they found $3.63 \pm 0.64 \ \textnormal{keV} / \textnormal{cm}^2 / \textnormal{s} / \textnormal{sr}$ (after translating to our units).  The masking removes bright AGN, but faint AGN could remain.  It will also remove the more massive halos, and so a direct comparison to our predictions is difficult without a precise understanding of the selection function.

\subsection{Cross-correlation of X-ray and tSZ}

\begin{figure*}
\includegraphics[width=0.70\textwidth]{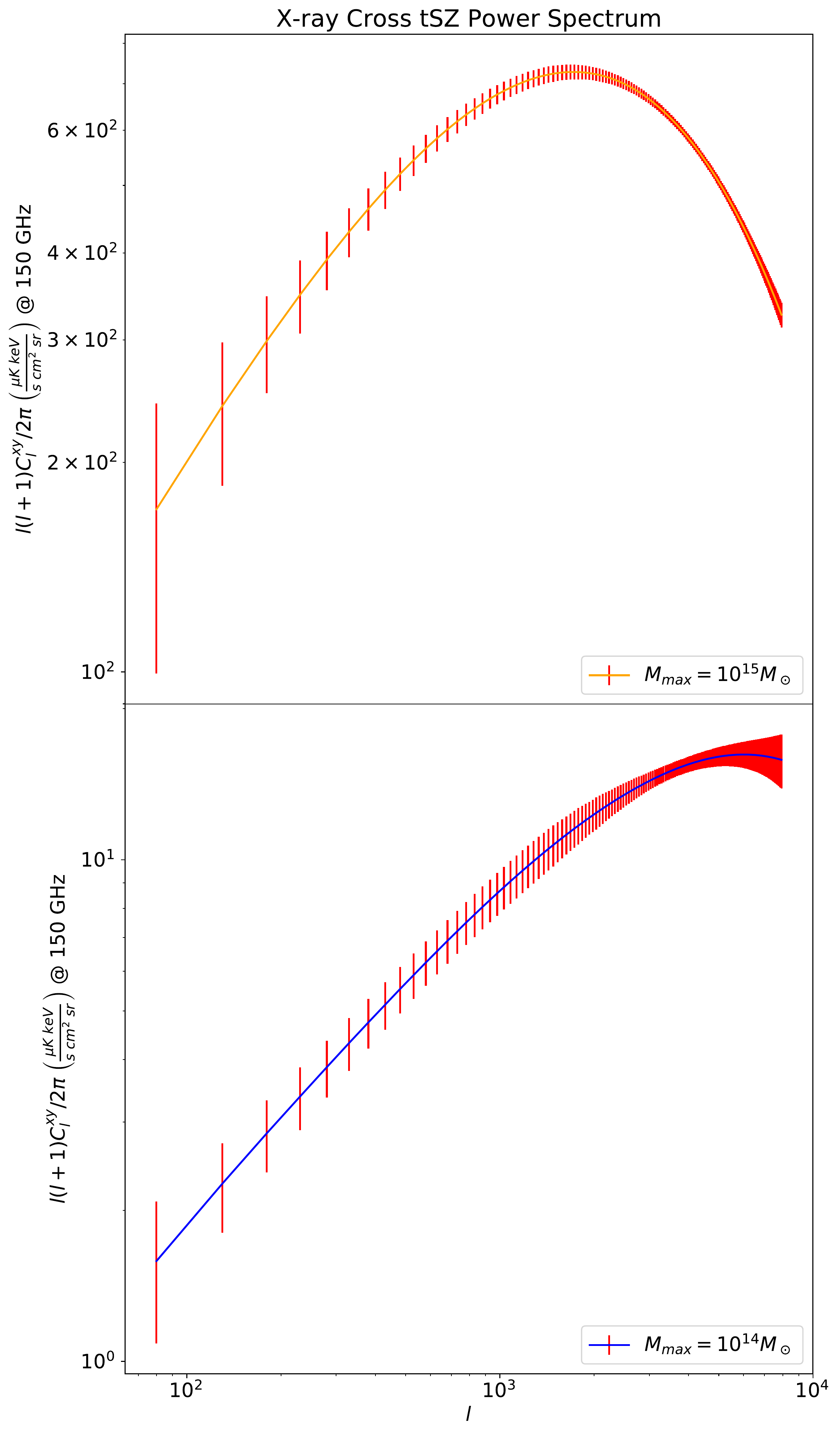}
\caption{X-ray cross tSZ power spectrum for clusters, groups and galaxies with error bars ($\Delta l = 50$) for eROSITA cross-correlated with the Simons Observatory. Since the relative values for the two- and four-point errors change for each mass cut, the error bars blow up at different multipoles. For $M < 10^{15} M_\odot$ halos, the overall SNR is 174. For $M < 10^{14} M_\odot$ halos, the signal is less, but the non-Gaussian contribution to the noise is proportionally lower, and the overall SNR increases to 201. For $M < 10^{13} M_\odot$ halos, the overall SNR is just 0.21, so we do not show it.}
\label{Cxy}
\end{figure*}

\begin{center}
\begin{figure}
\includegraphics[width=\linewidth]{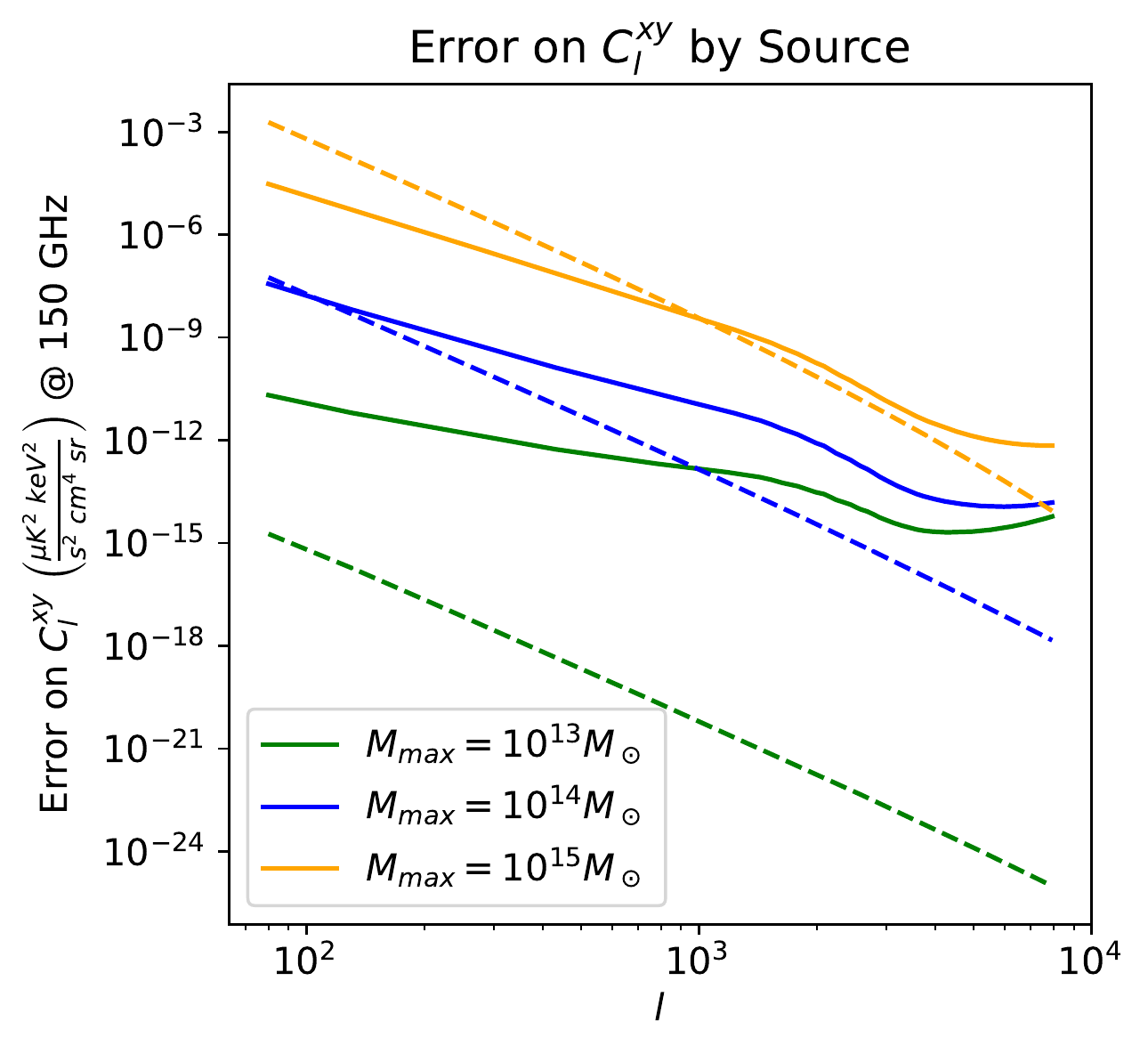}
\caption{Contribution to the uncertainty by $\Delta C^{xy}_l$ (solid line) and the diagonal of $T^{xy}_{ll'}$ (dotted line). The cluster noise is dominated by the four-point term until beam effects take over. For smaller halos, the dominant term switches in the lead up to the signal's peak. For galaxy-size halos, the two-point term is always dominant and becomes beam-effect dominated well before the signal's peak.}
\label{Cov}
\end{figure}
\end{center}

We show the X-ray--tSZ cross-power spectrum in Figure \ref{Cxy} with our forecast for uncertainties.  These experiments should give us a clear picture of much of the cross spectrum: we will be able to measure both the overall power and detailed shape, with an order of magnitude better signal-to-noise than what is currently possible with Planck and ROSAT. The uncertainties are largest at large scales (low multipole $l$), because of sample variance and the small number of modes, and small scales (high $l$), because the small-scale signal is suppressed by the instrument beam.  When all halos are included, the total SNR for the cross spectrum is 174. 

Because a substantial portion of the signal comes from rare, massive objects, the non-Gaussian (4-point) contribution to the errors due to (Poisson) sample variance is significant.  Figure~\ref{Cov} shows the various contribution to the errors.    Lower mass halos are more numerous, so by central-limit-theorem arguments, the errors are more 
Gaussian.  For this reason, we find that we can improve the signal to noise of the cross-spectrum by masking the most massive halos.  This effect is also seen in the tSZ auto power spectrum \citep{Hill2013}.  For example, masking halos down to mass $M<10^{14} M_\odot$ (a few times below the level already achieved by ACT and SPT) lowers the signal, but lowers the sample variance even more, and so raises the total SNR to 201. However, in order to obtain this result with real data, we would require extensive knowledge of the cluster selection function.  Otherwise, this procedure could lead to bias in the interpretation of the power spectrum that nullifies the usefulness of the increased SNR.

Such cluster finding and masking techniques already exist. \cite{SO2018} forecast the detection of all objects with $M > 1.5 \times 10^{14} M_\odot$ in a nearly redshift independent way.   \citet{Pillepich2012} compute that eROSITA could detect objects objects of $M > 10^{14}\  M_\odot$ with 50 or more counts for $z < 0.4$ and objects of $M > 10^{13}\  M_\odot$ for $z<0.1$.  At lower masses, the signal falls fast.  For comparison, if we were able to mask halos to $M<10^{13} M_\odot$ for all redshifts, we could set limits on the overall amplitude of the cross-power with an expected SNR of 0.21.  

\subsection{Parameter Constraints}
We want to see what cosmological and gas physics information we can extract from our predicted cross-spectrum, and use a Fisher matrix analysis to guide us. Here we examine halos of $M < 10^{15} M_\odot$ for our Fisher analysis. The Fisher information matrix is defined as:

\begin{equation}
\mathcal{F}_{ij} = \sum_{ll'} \frac{\partial C_l^{xy}}{\partial \theta_i} \left(\sigma^2_{ll'}\right)^{-1} \frac{\partial C_l^{xy}}{\partial \theta_j},
\end{equation}
where $\theta_i$ and $\theta_j$ are parameters that affect our power spectrum and $\sigma^2$ is the cross-spectrum covariance given in Equation \ref{sigmasquared}.
We take the inverse of the Fisher information matrix to get the parameter covariance:
\begin{equation}
{\rm Cov}(\theta_i,\theta_j) = \left( \mathcal{F}_{ij} \right)^{-1}.
\end{equation}
The diagonal elements of the covariance are the squares of the individual parameter errors. 

To normalize the vastly difference scales of our fitting coefficients, we scaled all the coefficients by their fiducial value
\begin{equation}
\theta_{i} = \frac{\theta_{i,\, \textnormal{original}}}{\theta_{i,\, \textnormal{fiducial}}}. 
\end{equation}

Our model has many fitting coefficients to consider (29 in total), and despite the high signal-to-noise measurement of the cross-power spectrum, we found that a full Fisher analysis yields no significant constraints on any one coefficients, marginalized over all the others.  This is due to degeneracies between the effects that fitting coefficients have on the cross spectrum.   Degenerate effects lead to linearly dependent rows in the Fisher matrix, which cannot be disentangled.  So in practice we will have to fix parameters to sensible values, provide priors on those parameters from other measurements or simulations, or combine into single effective parameters those parameters with similar effects.

The auto power spectra could contain additional information that may in principle resolve some of the degeneracy between our parameters.  The tSZ power spectrum is relatively clean, but the X-ray auto spectrum would suffer contamination from extragalactic, Milky Way, and Solar System sources.  This complicates its use.  The computation of the covariance between the auto power spectra and the cross power spectrum is more complicated than what we have presented here (particularly in the four-point noise terms) and so we have not yet included the information from the auto spectra.

We explore parameter constraints further in two ways.  First, we looked at the fitting coefficients one at a time to identify which are most important to the cross-spectrum (assuming other parameters are fixed, not marginalized), and find their response to added outside cosmological information.  Second, we looked jointly at subsets of the important coefficients to examine degeneracies.

For the single-coefficient constraints, we looked at two cases.  In the first case we fixed the cosmology to be known exactly and in the second case we allowed the cosmology to vary within priors set by future SO constraints. The fiducial values and priors for SO constraints are shown in Table \ref{table:CMBS4priors}. In both cases, we only allowed one gas physics coefficient to be a free parameter in the analysis and the rest were fixed to their (assumed known) fiducial values.  The results of this analysis are shown in Table \ref{1by1}. The perfectly-known-cosmology column gives raw sense for how much a parameter can be constrained by these measurement, ignoring degeneracies with other parameters. The upshot is that we are more sensitive to the overall scales for the pressure and electron density and less sensitive to the mass and redshift dependence.  Clumping coefficients tend to be more poorly constrained, with the mass-dependence of the core radius as the best-constrained parameter. 

When we examine single coefficient constraints in the second case, with a more realistic future cosmology prior, the constraints get worse but we keep the same trend.  The overall scales of pressure and gas density are the most important. From this analysis, we conclude that the best parameters to focus our efforts on are fitting scales of the electron density and pressure as well as the redshift dependence of the pressure's core radius and amplitude.

For joint constraints of subsets of coefficients, we took our list of the best-constrained coefficients and first examined all pairs of gas coefficients, and each gas coefficient paired with each cosmological parameter, to look at degeneracies between them.  To do this, we looked at two quantities in turn: the off-diagonal entries of the product between the 2-parameter Fisher matrix and its inverse (the product should be the identity), and the condition number of the covariance. If the off-diagonal entries are zero within numerical accuracy, we can say that our inversion of the Fisher matrix was successful and thus the condition number of the matrix gives us a quantitative assessment for the degeneracy between parameters.  

As the next step, we expanded our analysis and found that we could constrain no subset with six or more parameters: the condition numbers of the matrices were so high that no information could be obtained from attempting to disentangle the sets. 

From subsets of five and fewer parameters, we saw a trend in coefficient subsets that had low condition numbers.  The list of these more robust coefficients is given in Table \ref{multiparam}. These gas coefficients---in addition to cosmological parameters $\Omega_M$, $\Omega_B$, and $n_s$---in combinations provided the best conditioned matrices, better by an order of magnitude compared to the other gas coefficients. By being well-conditioned, these coefficients are not degenerate and can be studied independently.  From the multiparameter analysis, we conclude that the overall scales of the pressure and electron density profiles and the redshift dependence of the pressure's amplitude and core radius are the parameters we learn the most about from this kind of analysis. In order to learn more about redshift dependence, mass dependence, and clumping, we will need other types of measurements.

\begin{table}
\centering
\begin{tabular}{||c c c||} 
 \hline
 Parameter & Fiducial Value & Prior \\ [0.5ex] 
 \hline\hline
 $H_0$ & $69.0$ & $0.24$ \\ 
 $\Omega_M$ & $0.2987$ & $0.0013$ \\
 $\sigma_8$ & $0.831$ & $0.014$ \\
 $N_s$ & $0.966$ & $0.002$ \\
 $\Omega_B$ & $4.7 \times 10^{-3}$ & $6.3 \times 10^{-5}$\\
\hline
\end{tabular}
\caption{SO projected cosmological parameters and prior constraints that we incorporate into our Fisher analysis.}
\label{table:CMBS4priors}
\end{table}

\begin{table}
\centering
\begin{tabular}{||c c c c c||}
\hline
Quant. & Param. & Coeff. & Perfect Cosmo. & SO Cosmo. \\ [0.5ex]
\hline \hline
$P$ & $x_c$ & $A_0$ & 0.002467	& 0.046616 \\
$P$ & $P_0$ & $A_0$	& 0.002683	& 0.098819\\
$P$ & $\beta$ & $A_0$	& 0.003497 &	0.044325\\
$n_e$ & $\alpha$ &$A_0$	& 0.003552 &	0.051189\\
$n_e$ & $\rho_0$ & $A_0$ & 0.006416 &	0.272122 \\
$n_e$ & $\beta$ & $A_0$	& 0.006457 &	0.067080\\
$P$ & $x_c$ & $\alpha_z$ & 0.007579 &	0.103171 \\
$P$ & $P_0$ & $\alpha_z$ & 0.009461 & 	0.095233\\
$n_e$ & $\rho_0$ & $\alpha_m$ & 0.012930 &	0.371041 \\
$P$ & $P_0$ & $\alpha_m$ & 0.014693 &	0.444804\\ 
$P$ & $\beta$ & $\alpha_z$ &0.017178 & 	0.216495 \\
$C_{2\rho}$ & $x_c$ & $q_2$ & 0.025299 &	0.107597\\
$n_e$ & $\rho_0$ & $\alpha_z$ & 0.025629 & 0.322954\\ 
$n_e$ & $\alpha$ & $\alpha_z$ & 0.044961 &	0.513990 \\
$C_{2\rho}$ & $\beta$ & $q_1$ & 0.057674 & 3.527459 \\
$n_e$ & $\alpha$ & $\alpha_m$ & 0.063592 &	0.834000 \\
$P$ & $\beta$ & $\alpha_m$ & 0.069107 & 	0.408097 \\
$C_{2\rho}$ & $\beta$ & $q_2$ & 0.080385 &	1.002552 \\
$n_e$ & $\beta$ & $\alpha_m$ & 0.097377 &	0.833283 \\ 
$C_{2\rho}$ & $\gamma$ & $q_2$ & 0.243343 & 1.132789 \\
$P$ & $x_c$ & $\alpha_m$ & 0.251711 & 	2.148814 \\
$C_{2\rho}$ & $x_c$ & $q_1$ & 0.322772 & 	3.949196\\
$n_e$ & $\beta$ & $\alpha_z$ & 0.682467 &	5.947609 \\
$C_{2\rho}$ & $\gamma$ & $q_1$ & 0.910562 &	6.823735 \\
\hline

\end{tabular}
\caption{Fractional constraints on gas fitting coefficients, examined one at a time in the Fisher analysis. The perfect cosmology column denotes the constraint for the coefficient if all other coefficients in the analysis are known exactly, including cosmological constants. The SO-like cosmology column denotes the constraint if a realistic SO-like cosmology is used instead of a fully known one. Both the perfect and real cosmo columns give a raw constraint for each coefficient in the our analysis. The scales of the pressure and electron density parameters are in general the best constrained by our analysis. The improvement from real cosmology to perfect cosmology indicates how sensitive a coefficient to improving cosmology.}
\label{1by1}
\end{table}

\begin{table}
\centering
\begin{tabular}{||c c c c c||}
\hline
Quantity & Param & Fit Coeff. \\ [0.5ex]
\hline \hline
$P$ & $P_0$ & $A_0$ \\
$P$ & $P_0$ & $\alpha_z$ \\
$P$ & $\beta$ & $A_0$ \\
$P$ & $\beta$ & $\alpha_z$ \\
$P$ & $x_c$ & $A_0$ \\
$P$ & $x_c$ & $\alpha_z$ \\
$n_e$ & $\alpha$ & $A_0$ \\
$n_e$ & $\beta$ & $A_0$ \\
$n_e$ & $\rho_0$ & $A_0$ \\
\hline
\end{tabular}
\caption{Across all joint multiparameter analysis, the fitting coefficients most common to coefficient subsets with the lowest average degeneracy, as assessed by the condition number. These parameters appear in combinations with each other and the cosmological parameters $\Omega_M$, $\Omega_B$, and $n_s$.}
\label{multiparam}
\end{table}

\section{Discussion and Conclusions}
\label{sec:discussion}
In this work we applied a halo model approach to the X-ray--tSZ cross-power spectrum with the intent to forecast upcoming measurements from eROSITA- and SO-like experiments. For our gas model, we used fitting functions for the pressure, electron density, and clumpiness of the gas from Battaglia et al. (\citeyear{Battaglia2011, Battaglia2014, Battaglia2016a}) to model how the gas properties vary with radius, mass, and redshift. In addition, we used APEC continuum and line emissivities to model the X-ray emission of the gas. This approach allows us to tailor to our emission modeling to very specific current and future needs. Finally, we modeled the number density of objects using a Tinker halo mass function.

With our halo model, we had two independent checks on the robustness of our results: the tSZ auto-power spectrum and the contribution to the mean X-ray background due to halo objects. We compared to the tSZ auto-power spectrum to the Planck 2013 Compton-$y$ map auto-power spectrum \citep{Planck2013}, finding good agreement, and to the mean Chandra X-ray background measurement of \cite{HM06}, finding the same order of magnitude, but with the caveat that this is not a precise one-to-one comparison.  This provides us with some confidence that our implementation of the gas profiles, mass function,  emission modeling, and other computations are working as expected.

For the X-ray--tSZ cross-spectrum, we forecast that the combination of eROSITA- and SO-like experiments will be able to provide a very high signal-to-noise measurements.  Including all halos ($M < 10^{15} M_\odot$), we found a total SNR of 174, about an order of magnitude better than what is possible today \citep{Hurier2014,Hurier2015}.  These experiments are so sensitive that much of the noise comes from sample variance.  Furthermore, the signals are substantially non-Gaussian, since massive halos are rare objects.  As a consequence, we find that masking massive halos increases the signal-to-noise in the measurement.  For halos of $M < 10^{14} M_\odot$, we see a theoretical total SNR of 201. The ability to identify and mask clusters on the sky at this mass range is already fairly robust.  If we could mask halos down to $M < 10^{13} M_\odot$, we would expect to see a total SNR of the correlation of 0.21, and be able to set upper limits to the residual signal.

Finally, we looked at what gas coefficients are best constrained by our analysis.  We determined that the overall scales for the pressure and electron density of the gas were the best constrained in general.   From our multiparameter analysis, we found that combinations of fit scales, the redshift dependence of the pressure's core radius and amplitude, and the cosmological parameters $\Omega_M$, $\Omega_B$, and $n_s$ provided the best conditioned and least degenerate sets of parameters to analyze. We could gain insight on the rest of the set and make predictions about mass scaling, clumping, and the other parameter redshift dependences due to their high condition numbers, and thus degeneracy, by understanding the independent coefficients in Table \ref{multiparam} better.

Looking ahead, this type of halo modeling is a powerful and versatile tool that has many applications for future measurements.  With our tool, we have a few avenues we could explore.  Potentially the most interesting is to explore what can be learned from the the X-ray line emission and experiments with high energy resolution like eROSITA, which has energy resolution between 50 and 157 eV \citep{Merloni2012}.  A tighter energy band can cut out much of the continuum emission and highlight the lines.  For example, the $0.4$--$0.6$ keV band can be sensitive to Oxygen lines that are excited in the WHIM, but not in higher-temperature gas, which allows us to study spatial correlations in the WHIM component and to forecast its detectability.  We could also calculate the correlations of tomographic slices of redshifting of X-ray lines.  Furthermore, we could replace either observable with other line of sight observables such as weak lensing, Cosmic Infrared Background emission \citep{CIB}, or the electron density measured by the dispersion relation of Fast Radio Bursts \citep{FRB} or the kinematic SZ effect \citep{Battaglia2017}.  Finally, we can look forward to CMB-S4 \citep{CMBS4} to provide further improvements to the noise level of the tSZ component, beginning in the late 2020s. With many upcoming experiments and the myriad of observables on the horizon, we will gain access to a plethora of information about the baryon component of the Universe.

\section*{Acknowledgements}
We would like to thank Nicholas Battaglia, Eugenio Ursino, Massimiliano Galleazzi, Nico Cappelluti, and Jack Hughes for discussions and input on the gas and X-ray modeling. We would like to thank Colin Hill for help applying the noise model for the Simons Observatory.  Although the authors participate in the Simons Observatory Collaboration, this work is not an official Simons Observatory paper.



\bibliographystyle{mnras}
\bibliography{library1}




\bsp	
\label{lastpage}
\end{document}